\documentclass[aps,prl,twocolumn,superscriptaddress]{revtex4-1}
\usepackage{graphicx}
\begin{document}
% Use the \preprint command to place your local institutional report
% number in the upper righthand corner of the title page in preprint mode.
% Multiple \preprint commands are allowed.
% Use the 'preprintnumbers' class option to override journal defaults
% to display numbers if necessary
%\preprint{}

%Title of paper
\title{Linearly dispersive bands at the onset of correlations in K$_x$C$_{60}$ films}

% repeat the \author .. \affiliation  etc. as needed
% \email, \thanks, \homepage, \altaffiliation all apply to the current
% author. Explanatory text should go in the []'s, actual e-mail
% address or url should go in the {}'s for \email and \homepage.
% Please use the appropriate macro foreach each type of information

% \affiliation command applies to all authors since the last
% \affiliation command. The \affiliation command should follow the
% other information
% \affiliation can be followed by \email, \homepage, \thanks as well.
\author{Ping Ai}
\affiliation{Materials Sciences Division, Lawrence Berkeley National Laboratory, Berkeley, California 94720, United States}
\author{Luca Moreschini}
\affiliation{Materials Sciences Division, Lawrence Berkeley National Laboratory, Berkeley, California 94720, United States}
\affiliation{Department of Physics, University of California, Berkeley, California 94720, United States}
\author{Ryo Mori}
\affiliation{Materials Sciences Division, Lawrence Berkeley National Laboratory, Berkeley, California 94720, United States}
\author{Drew W. Latzke}
\affiliation{Materials Sciences Division, Lawrence Berkeley National Laboratory, Berkeley, California 94720, United States}
\affiliation{Department of Physics, University of California, Berkeley, California 94720, United States}
\author{Jonathan D. Denlinger}
\affiliation{Advanced Light Source, Lawrence Berkeley National Laboratory, Berkeley, California 94720, United State}
\author{Alex Zettl}
\affiliation{Materials Sciences Division, Lawrence Berkeley National Laboratory, Berkeley, California 94720, United States}
\affiliation{Department of Physics, University of California, Berkeley, California 94720, United States}
\affiliation{Kavli Energy NanoSciences Institute at the University of California, Berkeley, and the Lawrence Berkeley National Laboratory, Berkeley, California 94720, United States}
\author{Claudia Ojeda-Aristizabal}
\affiliation{Department of Physics and Astronomy, California State University, Long Beach, California 90840, United States}
\author{Alessandra Lanzara}
\email[]{alanzara@lbl.gov}
\affiliation{Materials Sciences Division, Lawrence Berkeley National Laboratory, Berkeley, California 94720, United States}
\affiliation{Department of Physics, University of California, Berkeley, California 94720, United States}

%Collaboration name if desired (requires use of superscriptaddress
%option in \documentclass). \noaffiliation is required (may also be
%used with the \author command).
%\collaboration can be followed by \email, \homepage, \thanks as well.
%\collaboration{}
%\noaffiliation

\date{\today}

\begin{abstract}
Molecular crystals are a flexible platform to induce novel electronic phases. Due to the weak forces between molecules, intermolecular distances can be varied over relatively larger ranges than interatomic distances in atomic crystals. On the other hand, the hopping terms are generally small, which results in narrow bands, strong correlations and heavy electrons.
Here, by growing K$_x$C$_{60}$ fullerides on hexagonal layered Bi$_2$Se$_3$, we show that upon doping the series undergoes a Mott transition from a molecular insulator to a correlated metal, and an in-gap state evolves into highly dispersive Dirac-like fermions at half filling, where superconductivity occurs. 
%Therefore self-assembled molecular crystals provide a playground to realize quantum phases where highly dispersive electrons arise in the presence of strong electron interactions, alternative to the recently observed cases of twisted layer heterostructures.
This picture challenges the commonly accepted description of the low energy quasiparticles as appearing from a gradual electron doping of the conduction states, and suggests an intriguing parallel with the more famous family of the cuprate superconductors. 
More in general, it indicates that molecular crystals offer a viable route to engineer electron-electron interactions. 
\end{abstract}

% insert suggested keywords - APS authors don't need to do this
%\keywords{}

%\maketitle must follow title, authors, abstract, and keywords
\maketitle

% body of paper here - Use proper section commands
% References should be done using the \cite, \ref, and \label commands
%\section{}
% Put \label in argument of \section for cross-referencing
%\section{\label{}}
%\subsection{}
%\subsubsection{}
\newpage

Molecular assemblies are appealing systems for engineering correlations since, intermolecular forces being typically weak, they can often be patterned in a desired way by an appropriate choice of the substrate. As a notable example, C$_{60}$ can be grown on a number of substrates, resulting in different lattice mismatch, moir\'e periodicities, rotational order and contact geometries between the C$_{60}$ molecules \cite{Kiguchi_2003_ASS,LLWang_2004_rotation,Kim2015,Claudia_ACS_2017, Haag_2020_PRB}. Such flexibility allows acting on the balance between the on-site Coulomb repulsion $U$ and the bandwidth $W$, thereby tuning correlations and driving qualitatively different ground states. In this sense, self-assembling molecular crystals can represent a valid alternative to the recently discovered twisted-layer structures of graphene or transition metal dichalcogenides, \cite{Cao_2018_1,Cao_2018_2,ugeda2014giant,Wang2020,Xu2021}. Although these have the quality of lending themselves naturally to gating experiments, the ability to select a given ground state relies on the precise control of the twist angle, which is difficult and also incompatible with any bottom-up fabrication technique, since that would necessarily privilege the rotational alignment of the layers.

\begin{figure*}[t]
\begin{center}
\includegraphics [width=2\columnwidth,angle=0]{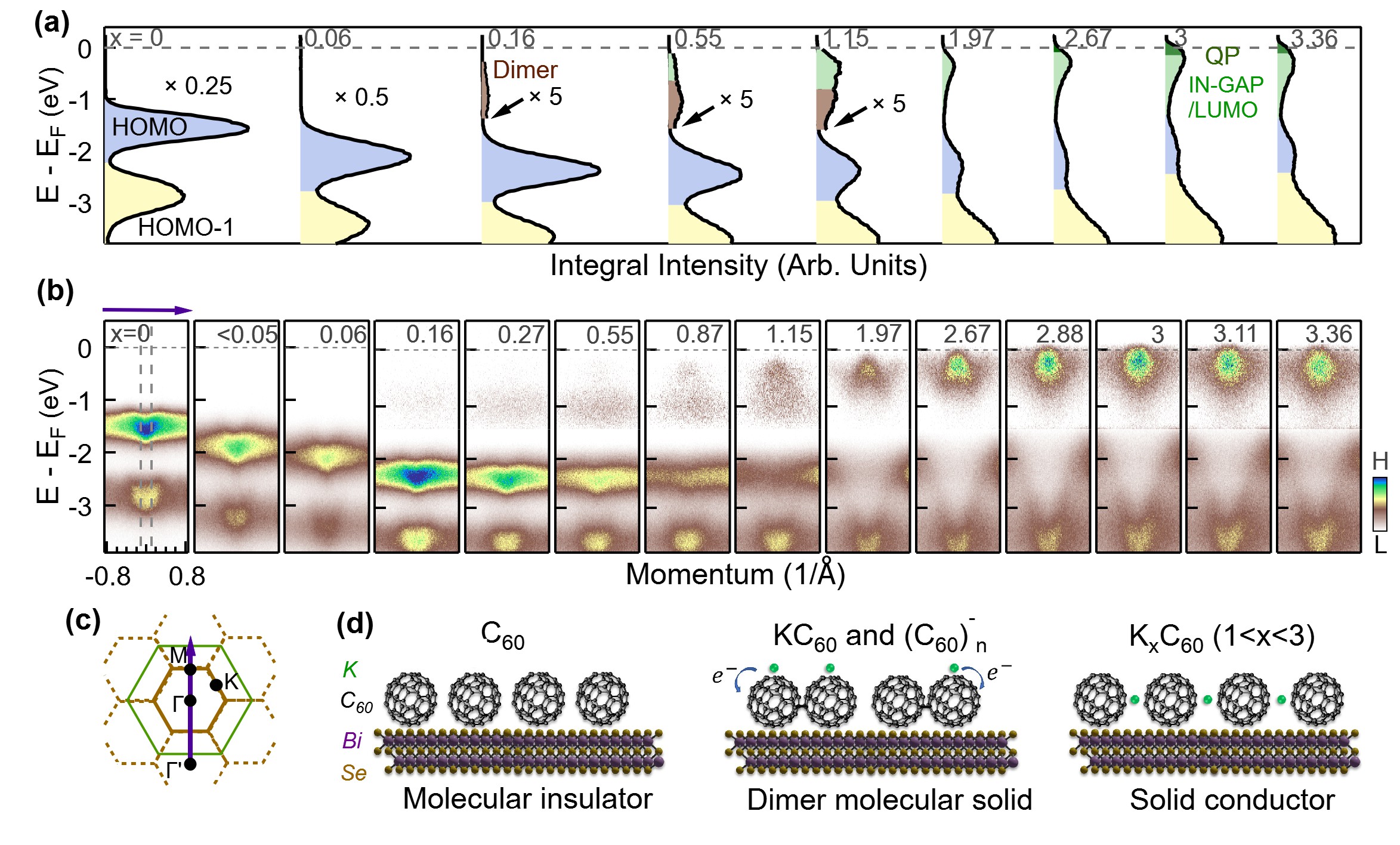}
%I have set the figure as the size of one/two column size, no need to resize it here, so the marks and texts in the figure will be all the same.
\end{center}
\caption {(a) Photoemission intensity integrated over a narrow range at $\Gamma$, indicated by dashed lines in the left panel of (b). The doping value is indicated at the top of each panel and is estimated from the intensity of the K 3$p$ levels, as explained in Ref.~\cite{supplinfo}. Some curves are rescaled by the numbers indicated. The different colors indicate the approximate energy range of the electronic states described in the text. (b) ARPES dispersion along the $\Gamma\mathrm{M}$ direction, as marked in the BZ in (c). For $x>0.06$ the intensity above and below $E=-1.4$~eV is separately rescaled by a factor of 10 and 2, respectively, to enhance the signal for the low energy states. (c) The 1$^{st}$ (solid lines) and the 2$^{nd}$ BZ (dashed lines) of the sample. The green lines indicate the first BZ of Bi$_2$Se$_3$ for reference. (d) Pictorial illustration of the transition from a molecular insulator to a solid conductor with the intermediate formation of dimerized pairs.
For the distance between the C$_{60}$ molecules in the plane we assume 9.66~$\mathrm{\AA}$ \cite{LatzkeDW_2019_ACS}, which yields $\overline{\Gamma\mathrm{M}}$ $\simeq$ 0.32 $\mathrm{\AA}^{-1}$.
}
\label{fig1}
\end{figure*}

Here we focus on the family of the A$_x$C$_{60}$ fullerides, where fullerene molecules act as anions and other elements, often alkali metals, act as cations. The majority of the A$_x$C$_{60}$ compounds are known to transition from insulating to metallic for increasing $x$ as the C $t_{1u}$ molecular orbitals get progressively filled \cite{Gunnarsson_1997_review,Capone_2009_RMP,Takabayashi2016}.
The trivalent members are mostly metallic and superconducting, but can be antiferromagnetic insulators for large intermolecular distances. The proximity of superconductivity to a Mott insulating ground state suggested the possibility that its nature, which remains to this day uncertain, may not be that of conventional BCS theory and derive instead from electron-electron interactions, bearing similarities with the case of other unconventional superconductors \cite{Capone_2002_science,Durand2003,Capone_2009_RMP,Manini2006,Takabayashi2016,Steglich2016,Yamamoto_2021}.

Although the effect of electron doping by alkali metals has already been studied in many instances by photoemission, providing evidence of a clear departure from a rigid band shift \cite{ChenCT_1991_nature,Merkel_1993_PRB, Wertheim_1993_PRB}, the lack of angular resolution has left many fundamental unanswered questions on how the metallic state emerges, what the nature of the low energy excitations is, and how correlated states such as superconductivity develop from this metallic phase.
Part of the difficulty associated with revealing the mechanisms behind these open questions is that, similarly to what happens in other C-based structures, such as graphene \cite{McChesney2010,Hwang2012,Cao_2018_1}, the low energy states of K$_3$C$_{60}$ appear to be extremely dependent on small modifications of the crystal structure, substrate or preparation conditions. In addition, the growth of A$_x$C$_{60}$ samples is known to be prone to phase separation \cite{Kawasaki2013,Brouet2006}. Indeed, the sparse photoemission measurements available present a remarkable variety of results both for angle-integrated \cite{Hesper_2000_PRB} and angle-resolved photoemission (ARPES) studies, spanning from a clear hole-like dispersion reported in films grown on Ag(111) \cite{YangWL_2003_science}, electron-like states on Ag(100) \cite{Brouet_2004_PRL} to a very shallow electron pocket on Cu(111) \cite{Pai_2010_PRL}.

Here, by using ARPES on \textit{in situ} grown films, we found that when doped C$_{60}$ molecules are deposited on top of the layered topological insulator Bi$_2$Se$_3$, a molecular subband develops, equivalent to a lower Hubbard band in a Mott scenario, accompanied by the evolution from massive to nearly massless quasiparticles as half filling is approached. These results place K$_3$C$_{60}$ in the regime of correlated metals, and point to the A$_x$C$_{60}$ series as a platform where highly dispersive, Dirac-like electrons coexist with, and possibly can be tuned or driven by, electron-electron interactions.

The samples measured in this work are K$_x$C$_{60}$ films with a thickness of five monolayers grown on Bi$_2$Se$_3$(0001), as described in the methods section \cite{supplinfo} and in Ref.~\cite{LatzkeDW_2019_PRB}. The C$_{60}$ molecules arrange along the (111) orientation in the same \textit{fcc} structure as in the bulk, but with a $\sim$3.4\% compressed lattice. Upon doping, the alkali ions progressively fill in the spaces in between the buckyballs until, for $x=3$, they form three hexagonal lattice layers for each C$_{60}$ layer, with $abab$ stacking \cite{Stephens1991}. The dispersion of the electronic bands remains mostly confined within the (111) planes (Supplementary Fig.S3)) and therefore we will discuss the data referring to the hexagonal surface Brillouin zone (BZ) of Fig.~\ref{fig1}(c).

The evolution of the electronic structure of K$_x$C$_{60}$ from the insulating C$_{60}$ to slightly above the ``optimal'' metallic state at half filling ($x=3$) is shown as a series of angle-integrated photoemission spectra in Fig.~\ref{fig1}(a) and as ARPES maps along the $\Gamma\mathrm{M}$ direction, measured with $p$ polarization, in Fig.~\ref{fig1}(b). In the angle-integrated data, the two peaks at $x = 0$ are associated with the highest occupied molecular orbital (HOMO) and the next highest occupied molecular orbital (HOMO-1). Adding potassium, which ionizes and donates one charge per atom to the molecular solid \cite{Martins1992}, for low doping levels both the HOMO and HOMO-1 spectral features shift toward higher binding energy as expected for a standard rigid band shift description upon electron doping, as well established from previous work \cite{ChenCT_1991_nature,Merkel_1993_PRB, Wertheim_1993_PRB}. The binding energy of the states is irrelevant in this regime since it depends on the pinning of the Fermi level ($E_F$) in the $\sim$1.6 eV band gap, further widened by the poor screening of the photohole \cite{Wertheim_1993_PRB}.

For $x>0.16$, a new in-gap state develops $\sim$1 eV below the Fermi level (brown in Fig.~\ref{fig1}(a)). Its existence only over a finite doping range, $0.16<x<1.2$, and its separation of $\sim$1.2~eV from the HOMO band (measured as peak-to-peak distance) suggests that it be related to the partial formation of C$_{60}$ dimer pairs that can be stabilized below 270K and prevents a simple metallic state in K$_1$C$_{60}$ \cite{Zhu1995,Oszlanyi_1995_PRB, Pichler_1997_PRL, Macovez_2007_PRB, Konarev_2010_ANIE}. For  $x>0.5$, an additional in-gap state emerges at $\sim$-0.5eV (green in Fig.~\ref{fig1}(a)), and in parallel an overall redistribution of the spectral intensity occurs between the HOMO and HOMO-1 bands. As we will show later (Fig.~\ref{fig1}(b) and Fig.~\ref{fig3}), this hybridization of the two molecular HOMO and HOMO-1 localized bands into a highly dispersive band for $x>1.5$, much more dramatic than predicted by theory \cite{Erwin_1991_science}, marks the departure from a narrow bandwidth molecular orbital picture and the formation of a solid crystal as schematically illustrated in (Fig.~\ref{fig1}(d)). As the doping approaches $x=3$, a point which coincides with the minimum resistivity and superconductivity \cite{Merkel_1993_PRB,Hesper_2000_PRB}, the in-gap state spectral weight is maximum and it crosses the Fermi level.

\begin{figure}[tbp]
\begin{center}
\includegraphics [width=1\columnwidth,angle=0]{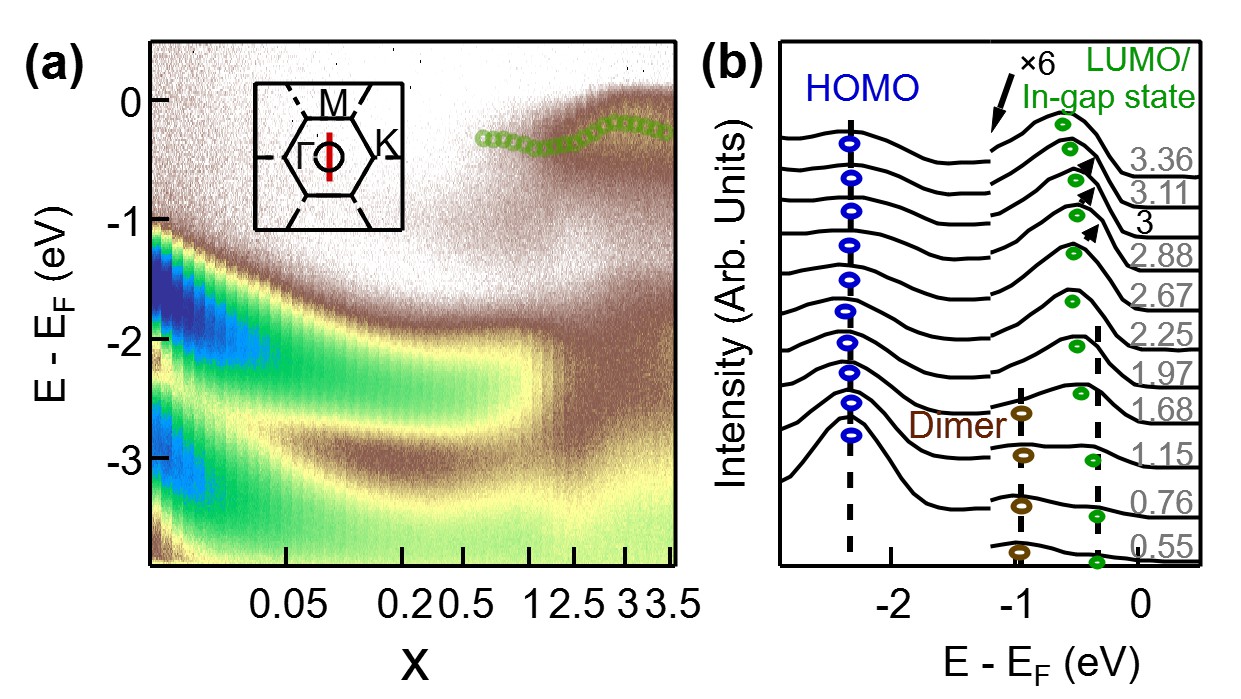}
\end{center}
\caption {(a) Detailed doping dependence of the ARPES spectra measured along ${\Gamma\mathrm{M}}$, as indicated in the inset, with $p$-polarized light. (b) A subset of the spectra zoomed in to the in-gap-state, dimer and HOMO states. The arrows for the spectra in proximity of $x=3$ point to the shoulder of the LUMO/in-gap state peak due to the metallic quasiparticle.
} 
\label{fig2}
\end{figure}

While the presence of both the insulating and metallic states has been previously reported, their origin has been matter of debate for a long time \cite{Gunnarsson1996,Han2000,Gunnarsson_1997_review}. Approaching the problem from band theory, the insulating phase for $x$=2,4 cannot be explained since the series is all expected to be metallic due to the partial filling of the $t_{1u}$ manifold. Conversely, starting from a Mott-Hubbard approach it is the metallic phase for $x$=3 that is equally puzzling, since the Coulomb repulsion $U$ is more than twice the $t_{1u}$ bandwidth $W$ ($\sim$0.5~eV) \cite{Satpathy1992}. To unambiguously understand how metallicity develops, it is critical to have access to the energy and momentum dependence of each state upon doping of the $t_{1u}$ manifold. We show in Fig.~\ref{fig1}(b) such a study for the first time. 

The data reveal a clear difference between the two in-gap states. Specifically, while the first one at $\sim$-1 eV is strongly localized, the second state in proximity of the Fermi level is clearly dispersive, and its band velocity appears to increase approaching half filling, while its binding energy decreases (as discussed in detail later, see Supplementary Fig. S6). This doping dependence supports an intriguing scenario where the near-$E_F$ states are associated with the formation of a lower Hubbard band driven by correlations and are responsible for the transition to the insulating phase of a solid crystal on both sides of half filling (K$_2$C$_{60}$ and K$_4$C$_{60}$). This sets the K$_x$C$_{60}$ fulleride series within a Mott transition framework on both sides of $x=3$, clearly different from the commonly accepted description where the low energy states are assigned to a lowest unoccupied molecular orbital (LUMO)-derived band, centered above $E_F$, which shifts into the occupied states as the doping increases, and the transition is interpreted in a band filling picture \cite{Merkel_1993_PRB,Wertheim_1993_PRB}. 

The doping dependence of the near-$E_F$ spectra is shown in Fig.~\ref{fig2} as an image plot of the ARPES intensity integrated over the momentum range along ${\Gamma}$M indicated by the red line in the inset. All the spectral features mentioned in Fig.~\ref{fig1} very rapidly shift to higher binding energy (note the non linear scale for the $x$ axis) and starting from $x\simeq1$ they start to gradually drift back toward $E_F$. 
The shift to higher binding energies pins the Fermi level to the bottom of the unoccupied LUMO states \cite{Merkel_1993_PRB, Wertheim_1993_PRB}, whereas the opposite shift (Fig.~\ref{fig2}(b)), not as clear in previous work and instead evident here, is almost certainly due to the more efficient screening approaching the metallic phase compared to that in the insulating one(s). 

At half filling, the quasiparticle weight also includes contributions from LUMO-derived states. In this regime the distinction between Hubbard bands and in-gap state is of course ill-defined as the states hybridize and close the gap \cite{Wang2020a}. The fact that this doping level also corresponds to the appearance of superconductivity represents an intriguing analogy with the case of cuprates, where the Fermi surface is best defined in the normal state at optimal doping, at the maximum of the superconducting dome.
%Note that while the presence of two subbands at either side of $E_F$ was directly visualized by scanning tunneling microscopy (STM) in K$_4$C$_{60}$ \cite{Wachowiak2005}, the electronic structure on the low doping side has not been equally well captured, despite the symmetry with respect to half filling reported by nuclear magnetic resonance, which finds for both the $n$=2 and $n$=4 occupancies a Mott-Jahn-Teller (JT) singlet ground state with a small spin gap  \cite{Brouet_2001_PRL,Brouet2002}. 

\begin{figure}[tbp]
\begin{center}
\includegraphics [width=1\columnwidth,angle=0]{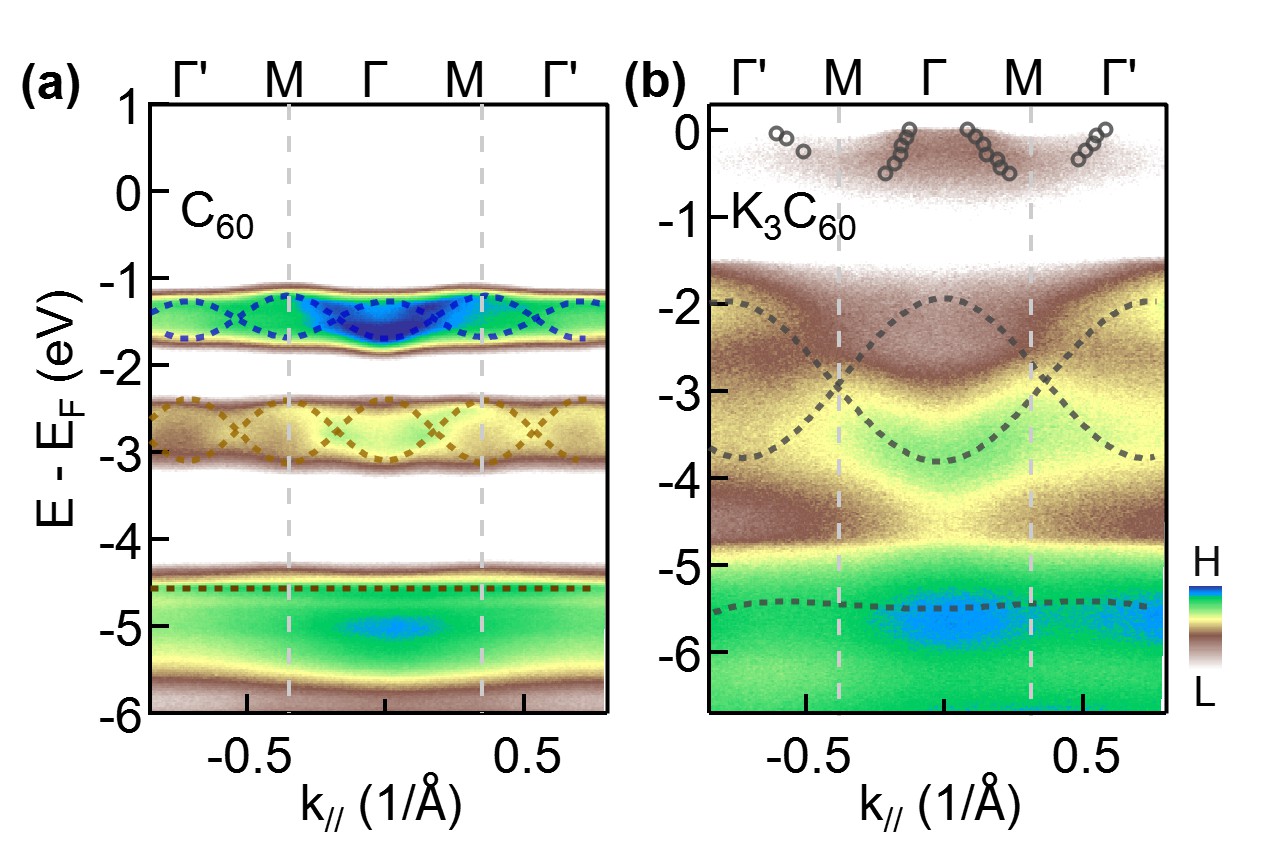}
\end{center}
\caption {(a) Band dispersion of undoped C$_{60}$ measured with $p$-polarized light. The dashed curves are guides to the eye following the dispersion of the HOMO states (for details see Ref.~\cite{LatzkeDW_2019_PRB}). (b) Same as (a) for K$_3$C$_{60}$. The symbols in the Fermi level region indicate the peak positions from MDC fits. 
%(c,d) Difference between left and right circularly polarized light ARPES maps for C$_{60}$ and K$_3$C$_{60}$. Note that the energy scale in (b, d) is shifted down for 0.7~eV relatively to that in (a, c) to align the HOMO/HOMO1 states.
Note that the energy scale in (b) is shifted by 0.7~eV relatively to that in (a) to align the HOMO/HOMO1 states.
} 
\label{fig3}
\end{figure}

In Fig.~\ref{fig3}(a,b) we compare the ARPES spectra along the $\Gamma\mathrm{M}$ direction at the two extremes of the undoped insulating phase of C$_{60}$ and of the optimally doped metallic phase of K$_3$C$_{60}$. As discussed in Fig.~\ref{fig1}(b), in the optimally doped phase, the merging of the C$_{60}$ molecular electronic levels with a bandwidth of $<$0.5~eV, into a single dispersive band with an overall bandwidth of $\sim$2~eV for K$_3$C$_{60}$ is observed. The K atoms therefore increase substantially the hopping terms between different fullerene buckyballs and the resulting bandwidth is now determined by the energy scale of the intermolecular forces \cite{Gunnarsson_1997_review}. Indeed, because of the hybridization between the two manifolds, the description of the orbital composition following the $I_h$ icosahedral symmetry and the strict separation between $\pi_5$ (H-type) character for the HOMO states and $\pi_4$ (G-type) for the HOMO-1 states (Ref.~\cite{LatzkeDW_2019_ACS}) is no longer valid.
Such hybridization is well visible also in the circular dichroism (CD) signal (See supplementary Fig.~S7).
%In K$_3$C$_{60}$ (Fig.~\ref{fig3}(d)) a single CD texture spans the whole bandwidth, supporting that the dispersion is associated with a single band, as opposed to undoped C$_{60}$ (Fig.~\ref{fig3}(c)), where the HOMO and HOMO-1 bands exhibit a clearly distinct CD signal \cite{LatzkeDW_2019_ACS}. The transition to delocalized states can be also indirectly inferred from the ARPES matrix elements. In K$_3$C$_{60}$, they nearly suppress the band maximum in the first BZ, yielding an apparent periodicity twice as large as the one of the unit cell, while the intensity distribution in C$_{60}$ is fairly even within the first two BZs.

At half filling an additional state can be extracted from the momentum distribution curve (MDC) fits, overlapped to the image (see markers in Fig.~\ref{fig3}(b)), which shows hole-like dispersion. At first sight it is reminiscent of the dispersive bands already reported for K$_3$C$_{60}$ grown on Ag(111) \cite{YangWL_2003_science,Brouet_2004_PRL}, yet a side by side comparison (Supplementary Fig.~S4) reveals that the two cases are distinct with a difference in Fermi velocity of over 70\% and a band minimum shallower than 0.2 eV in Ref.~\cite{YangWL_2003_science} \textit{vs.} a bandwidth larger than 0.4 eV in the present study.

\begin{figure*}[tbp]
\begin{center}
\includegraphics [width=2\columnwidth,angle=0]{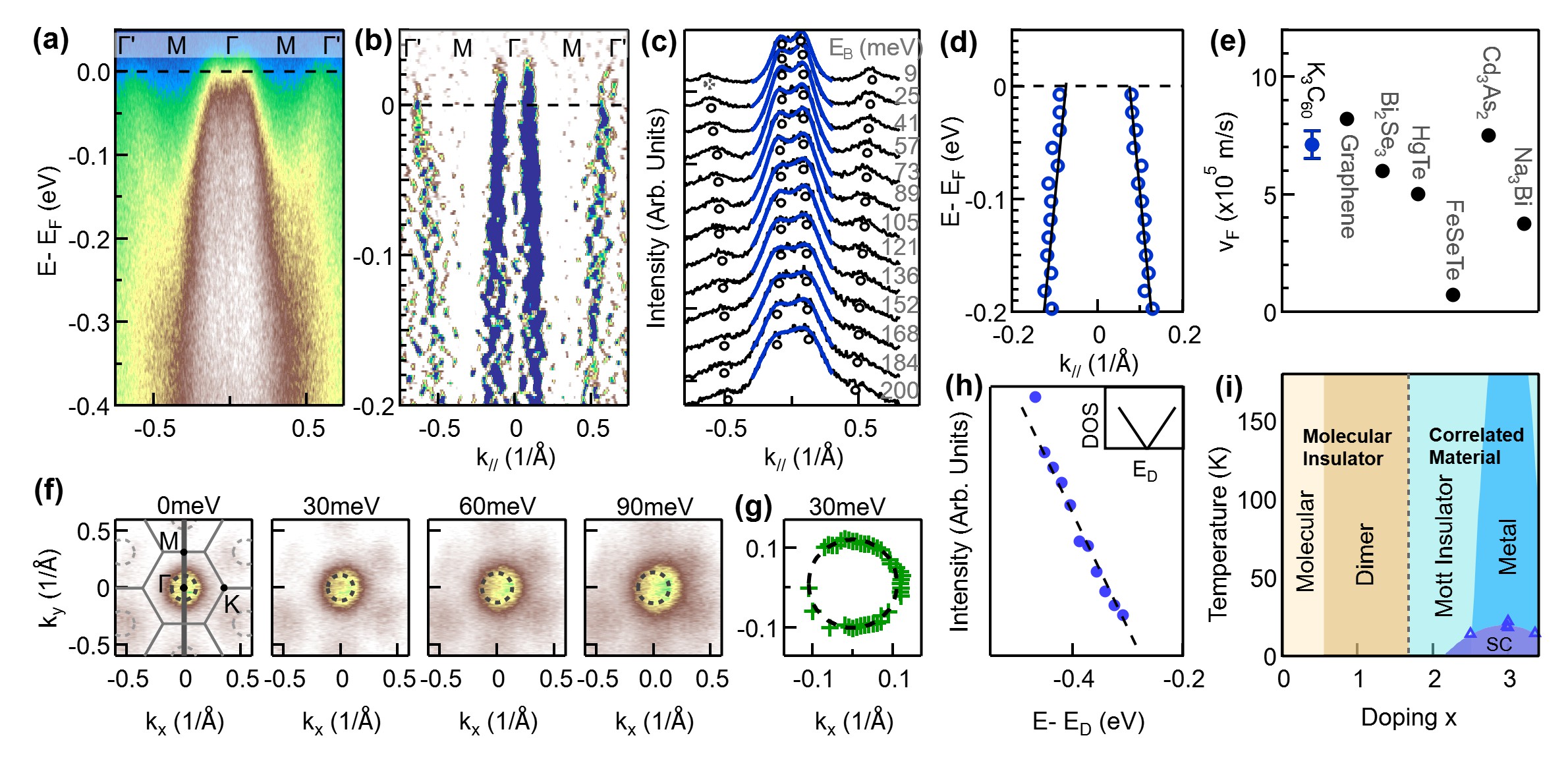}
%I have set the figure as the size of one/two columns size, no need to resize it here
\end{center}
\caption {(a,b) Low energy states of K$_3$C$_{60}$ along $\Gamma\mathrm{M}$ shown as an ARPES intensity map and a second derivative plot. (c) MDC stack in the same energy region, with the binding energy in meV indicated for each spectrum. The black curves are the ARPES data while the blue curves are best fits in a narrow energy interval around $\Gamma$ by a double Lorentzian function. The circles indicate the approximate peak positions. (d) The same circles reproduced together with a fit with a power function $E_D+A\cdot |x|^b$, where $E_D$ is the energy of the crossing point and the exponent $b$ is not constrained. The best fit was found for $b\simeq1.1$, very close to a linear function. The Fermi velocity from the closest linear fit is plotted in (e) together with that of several Dirac materials, and namely graphene \cite{Castro_2009_review, Hwang2012}, Bi$_2$Se$_3$ \cite{xia_2009observation}, HgTe \cite{Konig_2007_HgTe}, FeSe$_{0.45}$Te$_{0.55}$ \cite{Zhang2018}, Cd$_3$As$_2$ \cite{borisenko_2014experimental, neupane_2014observation} and Na$_3$Bi \cite{Liu_2014_Na3Bi}. (f) Constant energy contours at different binding energies up to 270 meV. The dashed circles in the left panel mimic the contours in the 2$^\mathrm{nd}$ BZs. (g) Peak positions from radial fits of the constant energy contour at 30 meV. (h) density of states as a function of the energy separation form the estimated crossing point, extracted from the integrated ARPES intensity. (i) Tentative phase diagram of K$_x$C$_{60}$ as a function of doping $x$. The blue triangles are from Ref.~\cite{QiqunXue_2020}, we did not investigate the presence of superconductivity in our films.}
\label{fig4}
\end{figure*}

%As mentioned in reference to Fig.~\ref{fig3}(e), the metallic state at half filling does not emerge from a downward shift of the LUMO states from above the Fermi level, but rather from an in-gap state shifting upward and merging with the LUMO states, in a Mott-transition-type fashion. This is shown more clearly in supplementary Fig.~S6 (c).

%where we plot the second derivative data taken along the momentum direction in the doping range $x\simeq 1.15 - 2.7$. In parallel with the upward shift for increasing doping, these clearly show an increase of the band velocity, as extracted from the fitting of the spectra along the momentum direction. Specifically, the results summarized in Fig.~S6(d) reveal a rather sudden jump, with more than 35\% increase at x$\sim$3, where correlations and superconductivity develop.

In Fig.~\ref{fig4}(a,b), we report a high statistics image measured along $\Gamma\mathrm{M}$ across two BZs for $x=3$. In contrast to the parabolic dispersion at lower doping, the data reveal a linear dispersion over the whole energy range up to 0.5 eV (see also supplementary Fig.~S4), and is reminiscent of massless Dirac fermions. The exact dispersion is extracted in a quantitative way by using a standard Lorentzian fitting of the momentum distribution curves in Fig.~\ref{fig4}(c,d). From a linear fit of the peak positions we find a Fermi velocity $v_F=(7.1\pm0.6)\times10^{5}\
$ ms$^{-1}$, comparable to some of the highest values reported for Dirac materials \cite{Castro_2009_review, Hwang2012, xia_2009observation, Konig_2007_HgTe, Zhang2018, borisenko_2014experimental, neupane_2014observation, Liu_2014_Na3Bi} (see Fig.~\ref{fig4}(e)).
%vf=slope/h=1/(6.5821*10^-16eV·s)*(xx eV·A)=xx/6.5821*10^16 A/s=xx/6.5821*10^6 m/s
The position of the crossing point can be determined by extrapolating the dispersion to the unoccupied states, and is estimated to be (0.3$\pm$0.1)~eV above the Fermi level. The constant energy maps ($k_x$ \textit{vs.} $k_y$), shown in Fig.~\ref{fig4}(f), as well as the radial MDC fits of Fig.~\ref{fig4}(g), prove the isotropic nature of the dispersion in the surface plane. Finally, the integrated spectral weight, which is proportional to the occupied density of states, as plotted in Fig.~\ref{fig4}(h), reveals the same linear dependence as a function of the binding energy characteristic of Dirac fermions \cite{Zhou_2006_NP}, in contrast to the more parabolic dispersion at lower doping. This description holds for a narrow doping range, with a transition to a parabolic hole-like dispersion for $1<x<2.7$ (Fig.~\ref{fig1}(b)).

Since fullerides are widely studied materials and can be grown on a number of substrates, we are left with educated guesses of the reasons why the linearly dispersive states at $\Gamma$ have not been reported before. The Bi$_2$Se$_3$ substrate used here is known to reduce the rotational disorder of the C$_{60}$ molecules \cite{LatzkeDW_2019_PRB}, and induces a compressive strain which lowers the intermolecular distance by $\sim$3.4\% with respect to the bulk value. This unusual situation (most substrates induce a tensile strain, see Supplementary Table S1) results in an increase of the bandwidth $W$, which is known to increase with the inverse of the intermolecular distance. Whereas on Ag(111) and Ag (100) values of $\sim$0.25 eV were found \cite{YangWL_2003_science, Brouet_2004_PRL,notePai}, here we extract a lower limit for $W$ of $0.4$ eV from the occupied portion of the $t_{1u}$ states, and up to $0.5-0.7$ eV depending on the (non accessible) dispersion above the Fermi level. The Coulomb repulsion $U$ on the other hand is less straightforward to estimate from ARPES and would require a more involved analysis and comparison with Auger spectroscopy \cite{Lof_1992_PRL}. A dedicated theoretical treatment would be welcome to determine if the increase of $W$ is accompanied here by a lowering of the $U/W$ ratio, which may be the culprit of the observed emergence of carriers of higher mobility at $\Gamma$. On the experimental side, a meaningful term of comparison will be provided by the recently synthesized fullerene crystals where the C$_{60}$ molecules are covalently bonded and the intermolecular distance is reduced with respect to that in the \textit{fcc} lattice \cite{Hou2022,Meirzadeh2023}.
%A fascinating possibility is to extend the present experiments to monolayer-thick films, which may host correlation-induced flat minibands, equivalent of the flat minibands observed directly by ARPES in twisted bilayer graphene \cite{Utama2020,Lisi2021}. In conjunction with massless fermions, these would provide the missing link to harness the potential of fullerides in the field of moir\'e superstructures, and serve as a molecular counterpart of atomic twisted heterostructures.

Regardless of the microscopic origin of the Dirac-like dispersion, the results presented reveal a transition from slow and massive electrons to fast and nearly massless quasiparticles which coexist and possibly are driven by electronic correlations. These findings put forward the exciting proposal that molecular crystals, with their adaptability to be patterned on multiple substrates, and in particular A$_3$C$_{60}$ compounds where even slight modifications of the bandwidth are known to result in different ground states, can provide a new flexible way to engineer correlations. In addition, the emergence of a metallic phase from an in-gap state in a fashion typical of a Mott transition hints at fascinating analogies with the case of superconducting cuprates. Interesting future developments will be to figure out how robust such massless fermions are in dependence of the intermolecular distance, as well as to explore the physics in the thin film limit, where the moir\'e potential would become important. K$_3$C$_{60}$ appears to be close to a sweet spot where the interplay between $U$ and $W$ allows for a crossover between an insulator and a correlated metal.

\begin{acknowledgments}
\section{Acknowledgments}
This work was primarily funded by the U.S. Department of Energy (DOE), Office of Science, Office of Basic Energy Sciences, Materials Sciences and Engineering Division under contract No. DE-AC02-05-CH11231 (Ultrafast Materials Science program KC2203). A.L. was partially supported as part of the Center for Novel Pathways to Quantum Coherence in Materials, an Energy Frontier Research Center funded by the US Department of Energy, Office of Science, Basic Energy Sciences. A.L. and L.M. also acknowledge support for sample growth from the Gordon and Betty Moore Foundation's EPiQS Initiative through Grant No. GBMF4859. C.O.-A. was funded for data acquisition by DOE Office of Science, Office of Basic Energy Sciences under Contract No. DE-SC0018154.  Sample growth was supported by the U.S. Department of Energy, Office of Science, Office of Basic Energy Sciences, Materials Sciences and Engineering Division under contract No. DE-AC02-05-CH11231, within the Van der Waals Heterostructures Program (KCWF16). The Advanced Light Source is supported by the DOE Office of Science User Facility under contract No. DEAC02-05CH11231.

\end{acknowledgments}

\section{Supplementary materials}
\subsection{Materials and methods}

%grow $C_{60}$ file
High-quality C$_{60}$ thin film samples were grown \textit{in situ} in ultrahigh vacuum on a bulk Bi$_2$Se$_3$ substrate as detailed elsewhere \cite{LatzkeDW_2019_PRB}. The film thickness was $\sim$50 \AA.
%deposition of K
Potassium deposition is done \textit{in situ} with a getter alkali evaporator. The angle-resolved and angle-integrated spectra are collected after each single deposition controlled by a timed shutter. 
The doping level is not linear with the dosing sequence and is estimated based on the relative ratio of the K 3$p$ core level intensity with respect to the optimum doping ($x = 3$) spectra.
The latter is identified by the maximum value of the ratio between LUMO and HOMO band, in line with previous reports \cite{Merkel_1993_PRB}). Note that, in view of our findings here, the LUMO intensity should be better identified as an in-gap (G) state, which hybridizes with the LUMO states only in proximity of half filling.
%ARPES measurement 
High resolution ARPES experiments were performed at Beamline 4.0.3 (MERLIN) of the Advanced Light Source using 45 eV linearly or circularly polarized photons, and 70 eV photons (in supplementary Fig.~S7) in vacuum better than $5 \times 10^{-11}$ Torr. The total-energy resolution was 20 meV with an angular resolution ($\Delta\theta$) of $\leq$ $0.2^{\circ}$. Data were taken at 200 K during K-dosing and then cooled to 30 K for the high statistics maps of the massless states. No obvious change was noticed lowering the temperature except for a trivial sharpening of the spectral features.

\setcounter{figure}{0}
\renewcommand{\thefigure}{S\arabic{figure}}
\begin{figure*}[t]
\begin{center}
\includegraphics [width=0.75\columnwidth,angle=0]{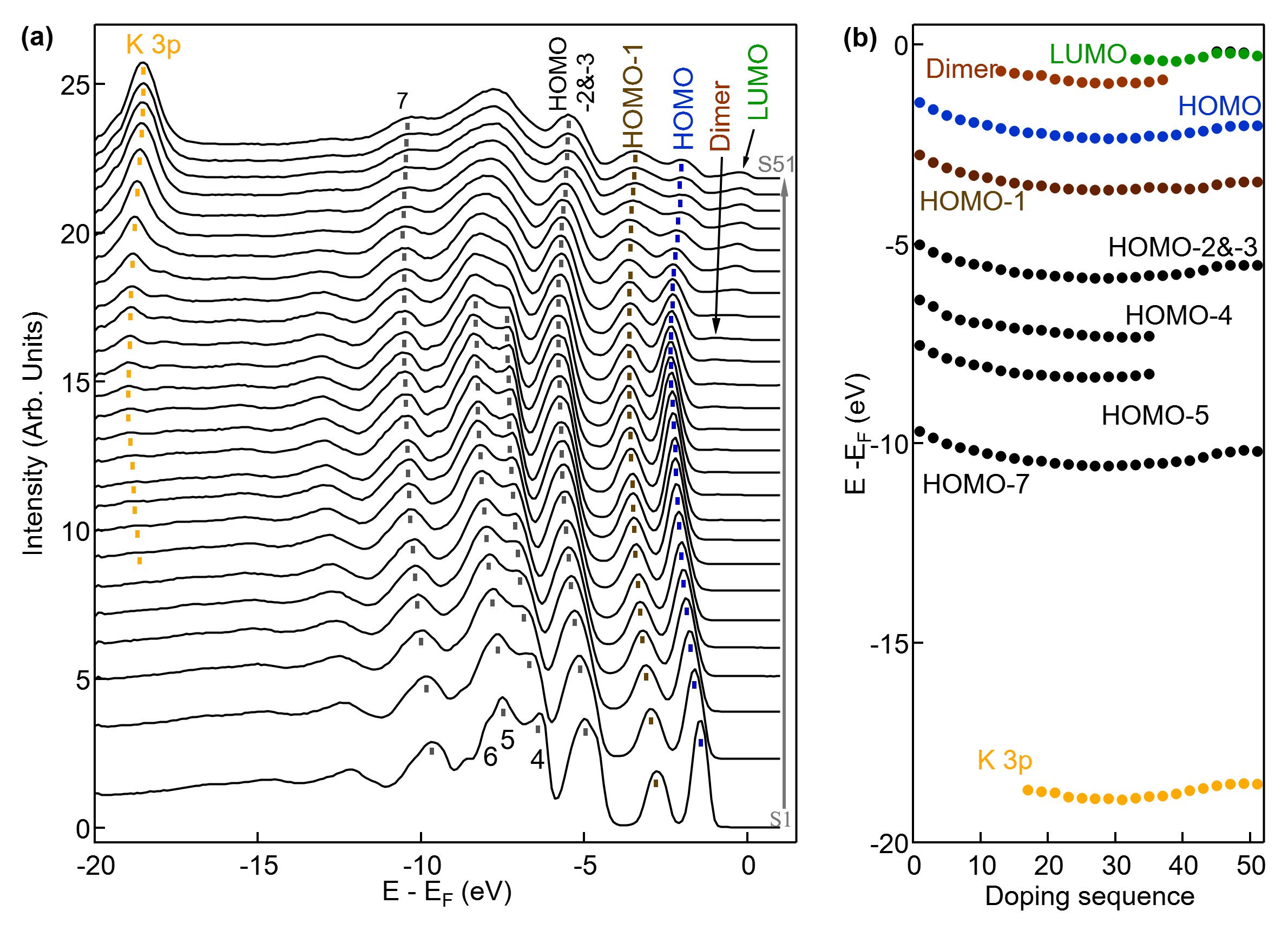}
\end{center}
\caption {\textbf{a} Angle-integrated photoemission spectra shown for increasing K doping. See the main paper for the assignment of the peaks at low binding energy. The peaks labeled with $4$-$7$ are assigned as common practice for the C$_{60}$ molecular levels\cite{Hasegawa1998,Kumar_2009_corelevel}. 
\textbf{b} Doping dependence of the peak binding energy of each occupied state from a fit of the spectra in \textbf{a}.}
\label{FigS1}
\end{figure*}

\vspace{3mm}
	
\subsection{Energy shifts upon electron doping}

In Fig.~\ref{FigS1} we show the evolution of the angle integrated photoemission spectra from undoped C$_{60}$ to a doping level of $x=3.36$. Overall, until the appearance of the G+LUMO states all the peaks including the K 3$p$ level follow a rigid shift, as observed in previous photoemission studies \cite{Merkel_1993_PRB, Wertheim_1993_PRB, Hesper_2000_PRB}. A more detailed dependence of the G+LUMO and HOMO energy shifts for doping levels approaching $x=3$ is shown in Fig.~2 of the main text.

Note that the energy shift changes sign as the sample starts to become more metallic, due to increased screening of the core hole, and that the HOMO and HOMO-1 peaks tend to merge, and same for the HOMO-4 and HOMO-5 features, as the addition of the K atoms increases the overlap between wavefunctions of different C$_{60}$ molecules and causes a departure from the molecular orbital description of the electronic levels.

%In Fig.~\ref{FigS1}(c,d) we plot the peak positions for all the doping values obtained and the angle integrated spectra for a subset of those, respectively, aligned to a common reference. This allows to better visualize the effect of correlations and the departure from a rigid band shift. We choose here the K 3$p$ level, when there is a clear peak, and the HOMO peak for the low doping range where no obvious K 3$p$ peak can be seen.
%The HOMO, HOMO-1 and the in-gap dimer state keep the same relative energy difference over the whole range, unlike the dispersive in-gap peak which follows a distinct trend, with a varying energy separation both from the higher binding energy states and from $E_F$. In proximity of the metallic phase, the spectra show an additional shoulder on the side of $E_F$, indicated by arrows in Fig.~\ref{FigS1}(d). This intensity is the quasiparticle signature of half filling, the spectral weight of which is strongly dependent on sample preparation conditions and photon energy \cite{Hesper_2000_PRB} and has remained elusive in some past photoemission studies \cite{Takahashi1992}.

\begin{figure*}[t]
\begin{center}
\includegraphics [width=0.5\columnwidth,angle=0]{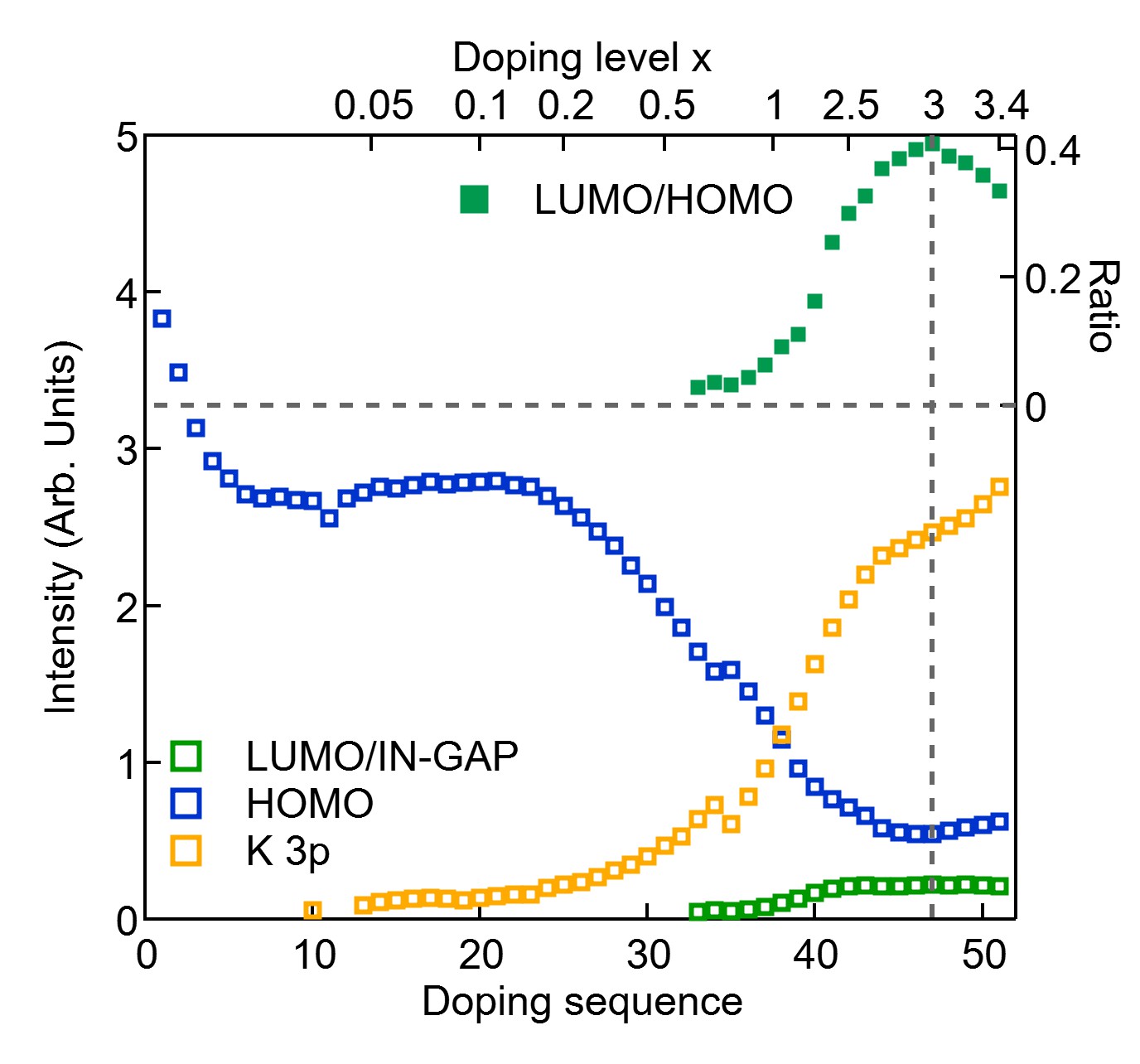}
\end{center}
\caption {\textbf{Doping level calibration.} The solid green squares represent the (G+LUMO)/HOMO intensity ratio. The point correspondent to the largest ratio is labeled $x=3$\cite{Merkel_1993_PRB}. Separate intensities for G+LUMO (green), HOMO (blue) and K 3$p$ (yellow) band are shown below. 
}
\label{FigS2}
\end{figure*}

\subsection{Doping level calibration}

Although it is known to vary as a function of the photon energy, the ratio between the photemission intensity of the G+LUMO and HOMO states in the K$_x$C$_{60}$ series is highest for $x=3$ \cite{Merkel_1993_PRB}. 
In Fig.~\ref{FigS2} we show separately the intensity of G+LUMO, HOMO and K 3$p$ levels extracted from the angle integrated photemission data, as well as the (G+LUMO)/HOMO ratio. The doping level is then calibrated relative to the maximum ratio. While this estimate has some margin of error, this is certainly lower than for a calibration based on the deposition time, since the alkali atoms may not distribute uniformly across the layers. Note also that the first deposition cycles result into a large shift of the electronic levels (see Fig.~\ref{FigS1}) even though the K 3$p$ is not yet visible in the spectra.

\vspace{3mm}
\subsection{Photoemission intensity oscillation of C$_{60}$ and K$_3$C$_{60}$}

\begin{figure*}[t]
\begin{center}
\includegraphics [width=0.8\columnwidth,angle=0]{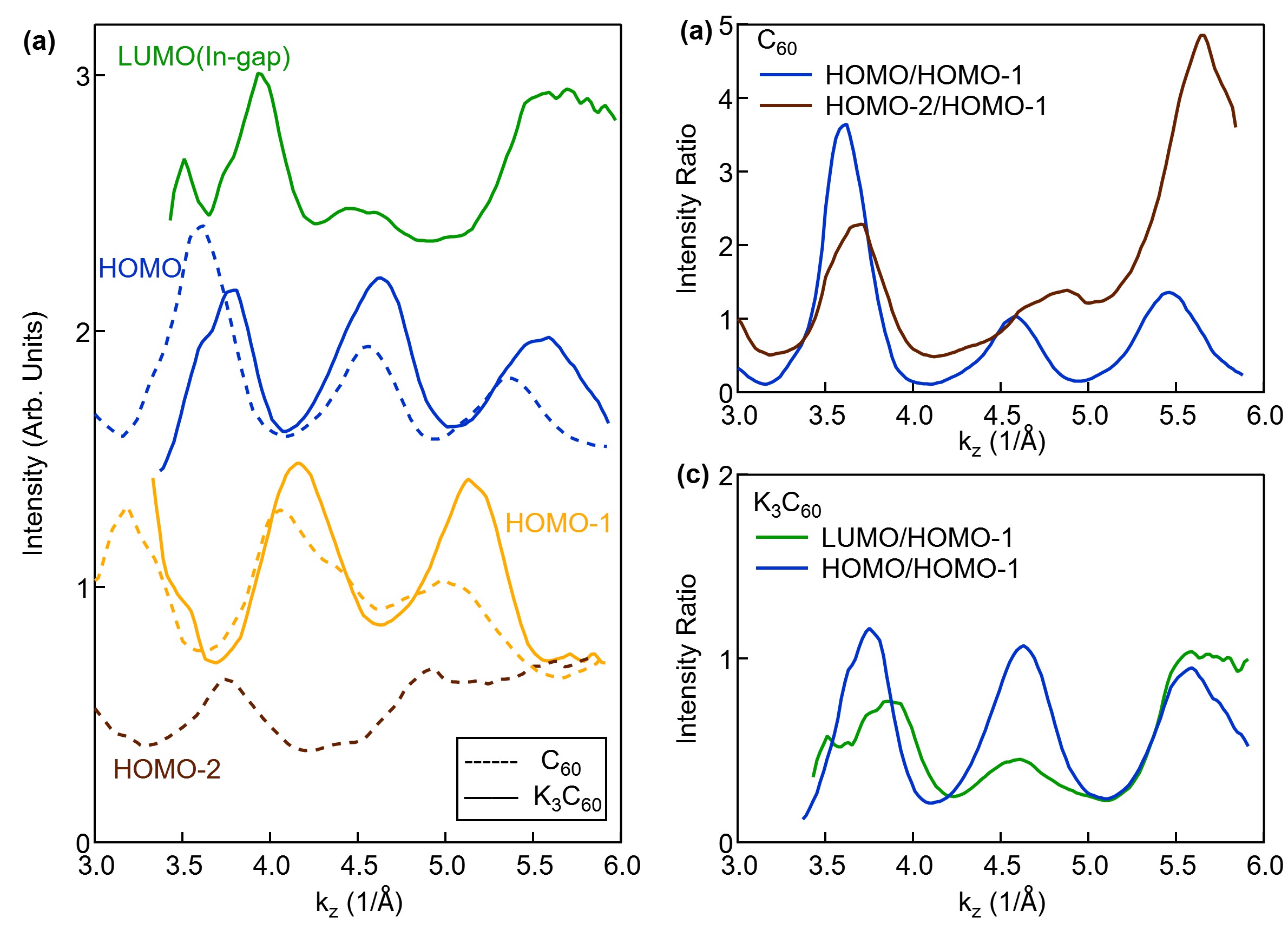}
\end{center}
\caption {\textbf{Photoemission intensity oscillation of C$_{60}$ and K$_3$C$_{60}$.} 
\textbf{a} The intensity oscillation G plus LUMO (green), HOMO (blue), HOMO-1 (yellow), and HOMO-2 (brown) from C$_{60}$ (dashed lines) and K$_3$C$_{60}$ (solid lines).
\textbf{b} shows the intensity ratio of HOMO/HOMO-1 and HOMO-2/HOMO-1 for normalization before doping. 
\textbf{c} shows the same for doped K$_3$C$_{60}$, but for (G+LUMO)/HOMO-1 and HOMO/HOMO-1.}
\label{FigS3}
\end{figure*}

The photon-energy dependences of photoemission intensities of C$_{60}$ were observed and discussed in previous papers \cite{Hasegawa_1998_oscillation, Andy_2002_oscillation, WANG_2008_oscillation, Daniele_2011_oscillation, LatzkeDW_2019_ACS}. Because of the unique molecular structure like a spherical shell and the large size of C$_{60}$, the photoemission intensity shows an oscillation for each molecular orbitals, as shown in Fig.~\ref{FigS3}. Density functional theory (DFT) and time dependent DFT level (TDDFT) calculations considering the the $\sigma-\pi$ symmetries of the ionized orbitals were carried out, which demonstrated the single-particle origin of the oscillations \cite{Daniele_2011_oscillation}.

\begin{figure*}[t]
\begin{center}
\includegraphics [width=1\columnwidth,angle=0]{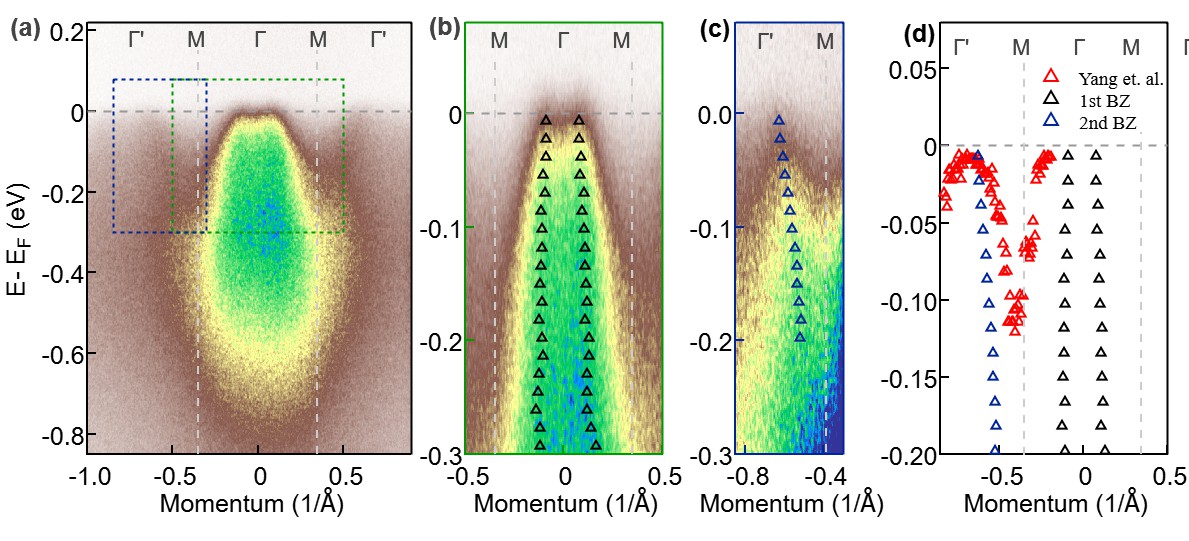}
\end{center}
\caption {\textbf{a} ARPES map along the $\Gamma\mathrm{M}$ high-symmetry direction, same as in Fig.~3 of the main text with larger energy and momentum range. \textbf{b} Close-up to the range indicated by the green dashed box in (a) (the 1$^{st}$ BZ) with fitted dispersion, same as in Fig.~3(a) of the main text.\textbf{c} As same as (b), but for the 2$^{nd}$ BZ (blue dashed box), with a different color contrast. \textbf{d} Fitted band dispersion from this work compared to the band dispersion from the previous Ref.~\cite{YangWL_2003_science}, shown as red triangles.
}
\label{FigS4}
\end{figure*}

\begin{figure*}[t]
\begin{center}
\includegraphics [width=0.9\columnwidth,angle=0]{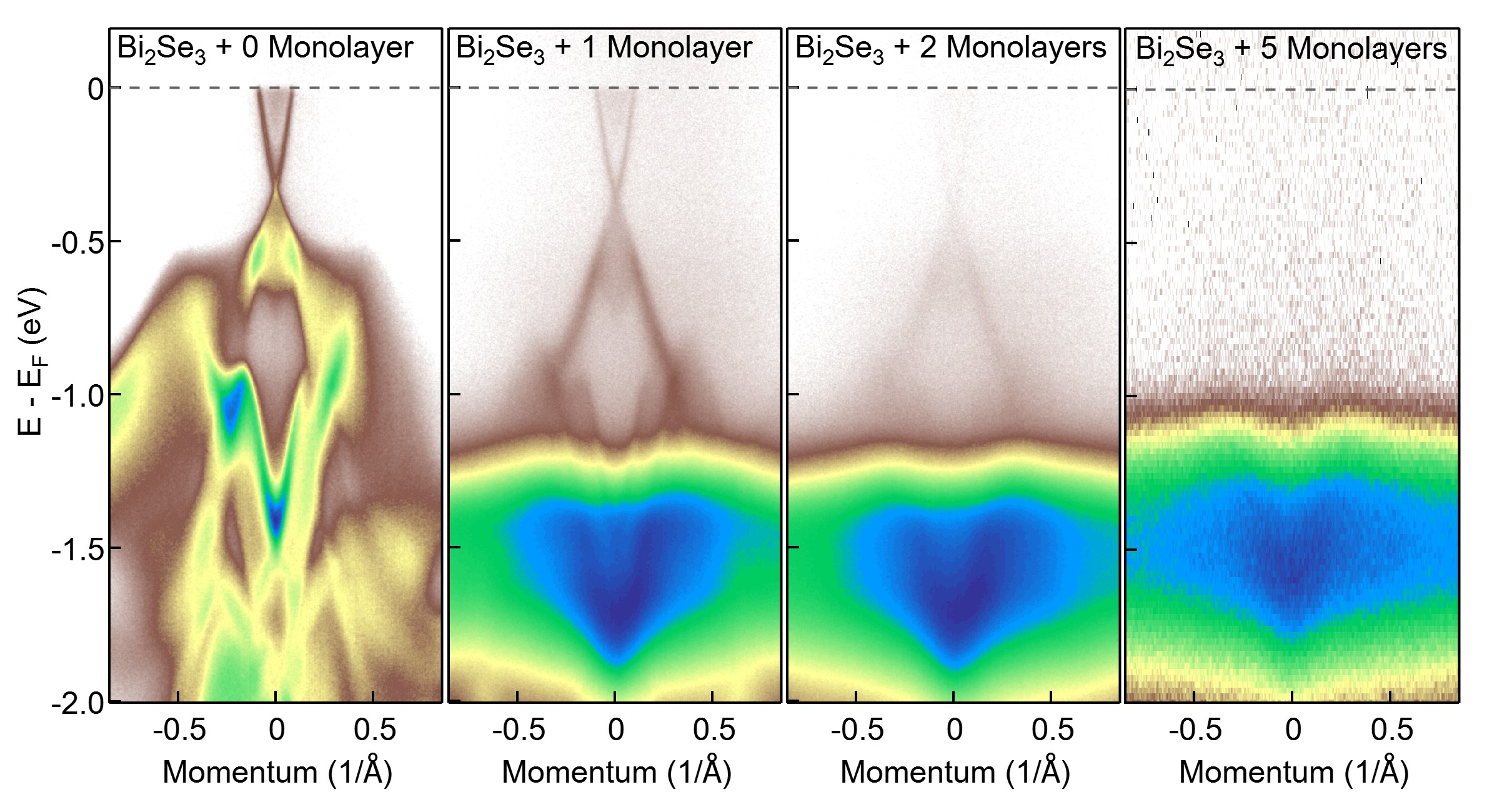}
\end{center}
\caption {From left to right, electronic structure measured along  
the $\Gamma\mathrm{M}$ high-symmetry direction for increasing C$_{60}$ coverage from 0 to 3 monolayers on a Bi$_2$Se$_3$(0001) substrate.
}
\label{FigS5}
\end{figure*}

Here we compare the dependence on the surface-normal momentum of the photoemission intensity for C$_{60}$ and K$_3$C$_{60}$ as total intensity (Fig.~\ref{FigS3}(a)) and as ratios of (G+LUMO)/HOMO, HOMO/HOMO-1 and HOMO-2/HOMO-1 (Fig.~\ref{FigS3}(b,c)) to help reduce the influence of the cross section variation. Clearly, although there are some differences among the overall intensity trends, the period of the $k_z$ oscillations is still dominated by the molecular properties of the C$_{60}$ molecule rather than by the period of the $fcc$ crystal structure, which would result instead in a $\sim0.78$ $\mathrm{\AA}^{-1}$ periodicity

\begin{figure*}[t]
\begin{center}
\includegraphics [width=0.9\columnwidth,angle=0]{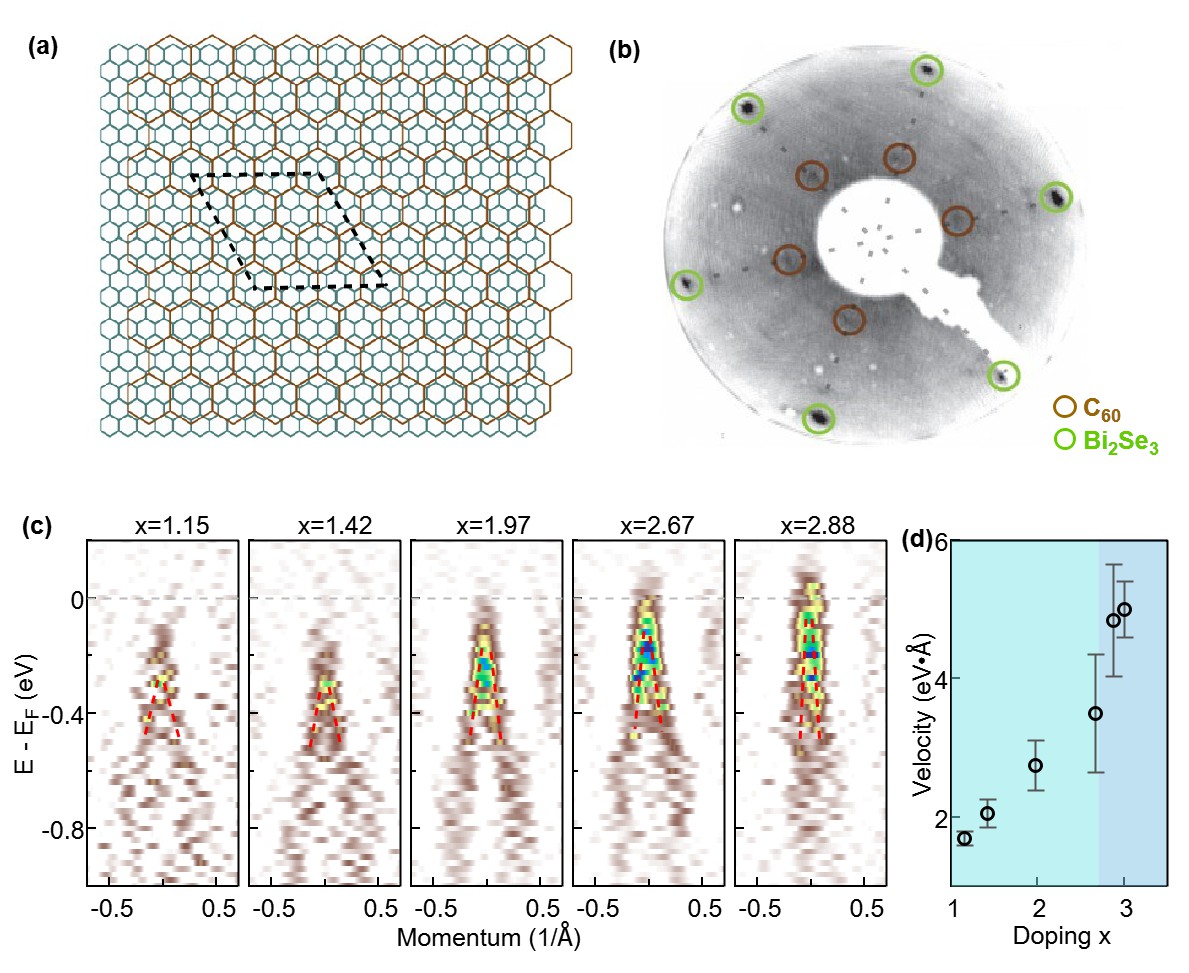}
\end{center}
\caption {\textbf{Band structure evolution in proximity of half filling.}
\textbf{a} Real space sketch of the C$_{60}$ balls on the Bi$_2$Se$_3$(0001) substrate (the additional atom (C$_{60}$ molecule) at the center of each small (large) hexagon is omitted for clarity). The rhomboids indicate the possible moir\'e pattern defined by the relative position of the C$_{60}$ ball and the Se atom. 
\textbf{b} LEED pattern measured at 34 eV for a single monolayer of C$_{60}$.
\textbf{c} Second derivative data taken along the momentum direction for the indicated doping levels. The dashed lines are guides to the eye.
\textbf{d} The band velocity extracted as $E=v\cdot k_{\parallel}$ assuming a linear dispersion for consistency across the doping range in panel \textbf{c}. The data point at $x = 3$ is taken from Fig.~4(d).
}
\label{FigS6}
\end{figure*}

\subsection{Linear dispersion at $\Gamma$}

\begin{figure*}[t]
\begin{center}
\includegraphics [width=0.6\columnwidth,angle=0]{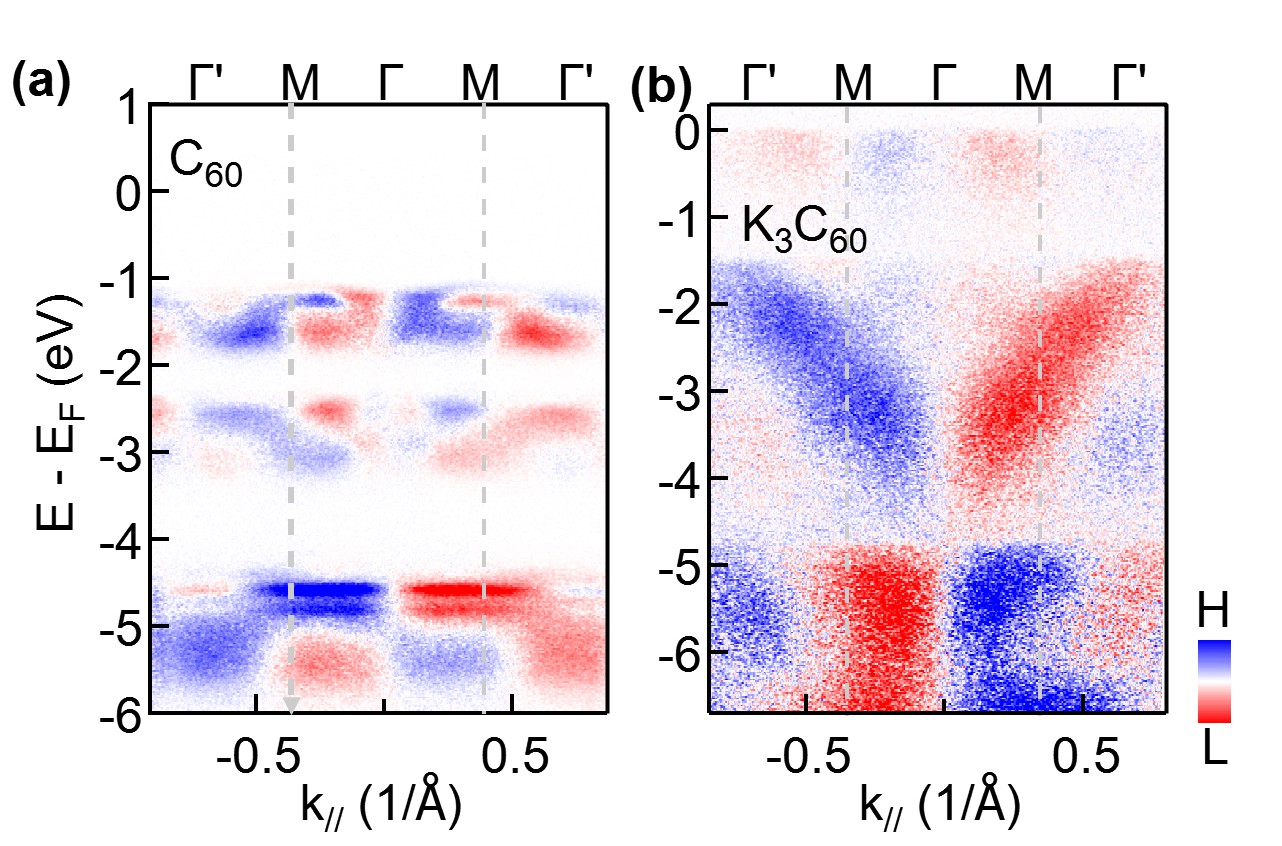}
\end{center}
\caption {(a) Difference between band dispersion of undoped C$_{60}$ measured with left and right circularly polarized light for C$_{60}$. (b) Same as (a) for K$_3$ C$_{60}$. 
Note that the energy scale in (b) is shifted down for 0.7~eV relatively to that in (a) to align the HOMO/HOMO1 states.
}
\label{FigS7}
\end{figure*}

Figure \ref{FigS4} shows the raw ARPES data on a larger energy scale than in the 2$^\mathrm{nd}$ derivative data of Fig.~3(a). The linear dispersion is better visible in the close-up in (b, c), where one branch at the side $\Gamma'$ points is also present at $|k_\parallel|\simeq0.7~\mathrm{\AA}^{-1}$. The bands are hole-like but have a qualitatively different dispersion than that reported in  Ref.~\cite{YangWL_2003_science} for K$_3$C$_{60}$ on a Ag(111) substrate, as evident in the comparison of (c). The latter data were taken on a single ML, even though the role of thickness in shaping the dispersion was not addressed. Extracting the Fermi velocity for the two cases we find $1.6\times10^{5}$ ms$^{-1}$
%0.7~eV$\cdot${\AA} 
for K$_3$C$_{60}$ in Ref.~\cite{YangWL_2003_science} \textit{vs.} $7.1\times10^{5}$ ms$^{-1}$
%4.68~eV$\cdot${\AA}
in the present study.

\subsection{Ruling out the presence of Dirac states from the substrate}

Figure \ref{FigS5} shows the Dirac cone of Bi$_2$Se$_3$ as its intensity fades upon increasing thickness of the C$_{60}$ film. Note that the crossing is in the occupied states at a very different energy than the one extracted for K$_3$C$_{60}$, and that the Dirac cone is completely suppressed for a thickness of 3 MLs, lower than the 5 MLs of the films studied here. Therefore the possibility that the low energy states would originate from the substrate can be ruled out.

\subsection{Crystal structure and in-gap band velocity}

Comparing the lattice constants of C$_{60}$ and Bi$_2$Se$_3$ we find that 3 times the C$_{60}$ intermolecular distance is equivalent to $\sim$7 times the in-plane Se-Se distance for Bi$_2$Se$_3$ (Fig.~\ref{FigS6}(a). This is consistent with the LEED pattern shown in Fig.~\ref{FigS6}(b) and with the ratio of the distances between the center and the reciprocal lattice points. Such reconstruction would mean for the C$_{60}$ films a 3.4$\%$ compression compared to the bulk $fcc$ value. This value is well within the range of the compressions/expansions observed on other substrates (see Table \ref{TableS1}) and therefore we consider this a reasonable assumption for an epitaxially grown film.

As mentioned in main manuscript, the band velocity increases with increasing doping. Here we plot second derivative data taken along the momentum direction in the doping range $x\simeq 1.15 - 2.88$, and extract the band velocity as $E=v\cdot k_{\parallel}$. We do so assuming a linear dispersion for consistency across the doping range in Fig.~\ref{FigS6}(c) to simplify the comparison, even tough the dispersion appearing linear at low doping levels is an artifact of the derivative along the momentum axis. Note that, as reported in numerous instances, the second derivative data are more reliable for highly dispersive features when the derivative is applied along the momentum direction, and viceversa for bands with small dispersion when the derivative is applied along the energy axis.
The results plotted in Fig.~\ref{FigS6}(d), along with an upward energy shift for increasing doping, clearly show an increase of the band velocity, and in particular a rather sudden jump, with a $>$$35$\% change, at x$\sim$3, where correlations and superconductivity develop.

\vspace{3mm}

\subsection{Circular dichroism of HOMO/HOMO-1 states}
As mentioned in the main text, the change in orbital character is also consistent with the circular dichroism (CD) signal from the HOMO/HOMO-1 states. In K$_3$C$_{60}$ (Fig.~\ref{FigS7}(b)) a single CD texture spans the whole bandwidth, supporting that the dispersion is associated with a single band, as opposed to undoped C$_{60}$ (Fig.~\ref{FigS7}(a)), where the HOMO and HOMO-1 bands exhibit a clearly distinct CD signal \cite{LatzkeDW_2019_ACS}. Note that the transition to delocalized states can be also indirectly inferred from the ARPES matrix elements of Fig.~3(b): in K$_3$C$_{60}$ they nearly suppress the band maximum in the first BZ, yielding an apparent periodicity twice as large as the one of the unit cell, while the intensity distribution in C$_{60}$ is fairly even within the first two BZs.

\vspace{3mm}

\subsection{C$_{60}$ growth on different substrates}

C$_{60}$ has been successfully grown on a number of substrates. In comparison with atomic crystals, for which films typically relax if exposed to values of epitaxial strain larger than 2-3\%, molecular crystals are more malleable and can be expanded or compressed by more than 5\%. Table \ref{TableS1} shows a list of the reconstructions reported so far on different substrates. In three cases the band structure was mapped by ARPES, with differing results: on Ag(111) the low energy states are hole-like \cite{YangWL_2003_science}, on Ag(100) they are electron-like \cite{Brouet_2004_PRL}, and same on Cu(111) but with a very different dispersion \cite{Pai_2010_PRL}. In this last case, as noted in the main text, we believe that the observed bands are due to an interface state.

Naively one would expect lattice compression to favor metallicity via an increase of the bandwitdh $W$, yet no obvious trend can be pointed out based on these sparse results since the effect on $U$ of the different reconstructions cannot be easily estimated.  
Note also that, out of the list in Table \ref{TableS1}, only Refs.~\cite{YangWL_2003_science}, \cite{Pb_table} and \cite{Au_table} are on monolayer-thick films. Whereas Ref.~\cite{YangWL_2003_science} finds a more standard $(2\sqrt3 \times 2\sqrt3)R30^{\circ}$ superstructure on Ag(111), confirmed later also in Ref.~\cite{Ag_table}, Ref.~\cite{Pb_table} reports on Pb(111) a large unit cell with a moir\'e wavelength $\lambda$ of 34{\AA} aligned with the C$_{60}$ lattice, and another of 46{\AA} with a twist angle of 11$^\circ$, and Ref.~\cite{Au_table} finds $\lambda\simeq70${\AA} 
on Au(111) and a rotational mismatch of 14.5$^\circ$. Based on the numerous reconstructions reported for thicker films it appears that the potential of fullerides for fabricating moir\'e heterostructures remains largely unexploited.

\newpage
\renewcommand{\thetable}{S\arabic{table}}  

\begin{table*}[tbp]
\resizebox{\textwidth}{!}{%
\begin{tabular}{cccccccc}
\hline \hline
substrate &
  Ag(111) &
  Al(111) &
  \multicolumn{2}{c}{Pt(111)} &
  Pb(111) &
  Au(111) \\
  \hline
structure &
  $2\sqrt3 \times 2\sqrt3 R30^{\circ}$ &
  { $2\sqrt3   \times 2\sqrt3 R30^{\circ}$} &
  \multicolumn{2}{c}{{ $\sqrt13   \times \sqrt13 R13.9^{\circ}$}} &
  HOC &
  { $2\sqrt3   \times 2\sqrt3 R30^{\circ}$} \\
intermolecular distance (\AA) &
  10.01 &
  9.91 &
  \multicolumn{2}{c}{10.005} &
  10.0 &
  9.99 \\
expansion(+)/compression(-) &
  1.1\% &
  -0.1\% & 
  \multicolumn{2}{c}{1.1\%} &
  1.0\% &
  -0.9\% \\
reference &
  \cite{YangWL_2003_science} &
  \cite{Al_table} &
  \multicolumn{2}{c}{\cite{Pt_table}} &
  \cite{Pb_table} &
  \cite{Au_table_R30} \\
  \hline
   &
   &
   &
   &
   &
   \\
   \\
  \hline
substrate &
  Cu(111) &
  Graphite &
  Si(111) &
  Si(100) &
  GaAs(001) &
  Au(111) \\
  \hline
structure &
  \begin{tabular}[c]{@{}c@{}}$\sqrt7   \times \sqrt7$ and $2 \times 2$\\ disordered and linear-wall maze \end{tabular} &
  Unknown &
  $7 \times 7$ &
  $2 \times 1$ &
  $2 \times 6$ &
  {$\sqrt589 \times \sqrt589 R14.5$}\\
intermolecular distance (\AA) &
  9.654 $\sim$10.224 &
  10.5 &
  unknown &
  10.4 &
  10.20 &
  10.02 \\
expansion(+)/compression(-) &
  -2.5 $\sim$ 3.3\% &
  6.1\% &
  unknown & 
  5.1\% &
  3.0\% &
  1.2\% \\
reference &
  \cite{Cu_table} &
  \cite{graphite_table} &
  \cite{Si7X7_table} &
  \cite{Si100_table} &
  \cite{GaAs_table} &
  \cite{Au_table} \\
  \hline
   &
   &
   &
   &
   &
   &
   \\
   \\
  \hline
substrate &
  Pt(111) &
  Cd(0001) &
  Cu(111) &
  \multicolumn{2}{c}{Ag(100)} &
  Bi$_2$Se$_3$(0001)\\
  \hline
structure &
  $2\sqrt3   \times 2\sqrt3 R30^{\circ}$ &
  Unknown &
  $4 \times 4$ &
  \multicolumn{2}{c}{c($6 \times 4$)} &
  $7 \times 7$\\
intermolecular distance (\AA) &
  9.5 &
  10.5 &
  10.2 &
  \multicolumn{2}{c}{10.42/11.56 (not hex)} &
  9.66\\
expansion(+)/compression(-) &
  -4.0\% &
  6.1\% &
  3.0\% &
  \multicolumn{2}{c}{4.2/5.6\%} &
  -3.4\% \\
reference &
  \cite{Pt_table2} &
  \cite{Cd_table} &
  \cite{Cu_table2} &
  \multicolumn{2}{c}{\cite{Brouet_2004_PRL}}&
  This work\\
  \hline
\end{tabular}
}
\caption{Reconstructions and periodicities reported for C$_{60}$ on several substrates. HOC = high order commensurate (multiple sizes and orientations).}
\label{TableS1}
\end{table*}

% Create the reference section using BibTeX:
%\bibliography{basename of .bib file}
\newpage

\bibliography{ref_K3C60}

\begin{thebibliography}{77}
\bibitem [{not()}]{notePai}%
  \BibitemOpen
  \href@noop {} {}\bibinfo {howpublished} {Due to the completely different
  fermiology, it is our best guess that the band structure reported for doped
  C$_{60}$ on Cu(111) in Ref. 29 be due to an interface state and largely
  unrelated to the K$_3$C$_{60}$ physics discussed here}\BibitemShut {NoStop}%
\end{thebibliography}
%\end{thebibliography}%

\end{document}